# Orientation of the spin axes of large Kuiper belt bodies as an indicator of the mechanism of their origin

Kazantsev A. M.
Astronomical Observatory of Taras Shevchenko National University of Kyiv
(Observatorna st., 3, Kyiv – 53, Ukraine, 04053)

ankaz@knu.ua

**Abstract**
The paper analyses the orientation of the spin axes of Kuiper Belt (KB) bodies that have satellites. For most of these bodies, the axes orientations were estimated under the assumption that the orbital planes of the satellites lie close to the bodies' equatorial planes. It was found that the spin axes of large scattered disks and plutinos' bodies, including four dwarf planets, are inclined at relatively small angles to the planes of their heliocentric orbits. This feature points to a possibility that all the scattered disk bodies, together with the bodies in the 3:2 resonance with Neptune, originated from the disruption of a single large body. The origin of the bodies of the so-called "cold" component of the KB is believed to differ from that of the scattered disk. The orientation of the spin axes of these bodies may also provide further evidence supporting this distinction.

## 1. Introduction

The possibility that the scattered disk (SD) objects in the Kuiper Belt arose from the destruction of a single large body was previously suggested by Kazantsev (2023). The destruction products of such a body (Kuiper Belt planet, KBP) could have formed most of the scattered disk objects in the KB, as well as objects in the 3:2 resonance with Neptune – plutinos. All four dwarf planets in the KB should also be considered products of this destruction.

The distributions of the *a–e* and *a–i* orbits of the SD bodies indicate the possibility that these bodies arose precisely as a result of the disruption of one large body. The semi-major axes of the orbits of the SD bodies reach values of $a >$ 500 AU. According to model calculations, for such values of $a$, the KBP fragment scattering velocities should have been greater than 2.0 km/s.

It is clear that such speeds, especially for large fragments, could not have been provided by the energy of the explosion at the moment of the KBP destruction. But parts of KBP could have had such speeds while rotating around the planet's axis. Then, when the KBP was destroyed, the fragments could have flown off at sufficient speeds.

According to Kazantsev's (2023) estimates, the KBP could have been close to Earth's size or larger. The destruction could have been caused by a collision with a large external body. Such a body could have originated from the cold KB component.

In addition, there are bodies hundreds of kilometres in size which, at perihelion, venture deep into Neptune’s orbit. At the same time, at aphelion, they reach distances exceeding 40 AU from the Sun. Such bodies are closer to the centaurs than to the scattered disk of TNOs (transneptunian objects). Some of these bodies may have transitioned to such orbits from the main asteroid belt (Kazantsev and Kazantseva, 2021).

If a not very monolithic structure of the matter have been inside KBP, a collision even with a noticeably smaller body could have led to its destruction.

The directions and propagation speeds of the fragments were determined mainly by the linear rotational velocity of these parts around the KBP axis at different depths and latitudes. The maximum propagation speed corresponded to the linear velocity at KBP surface near the equator and could have been 2.4 km/s.

The orbits of SD bodies have a characteristic distribution in the *a–i* coordinates. At the minimum values of *a* (about 42 AU), the inclination range is maximal and quite significant (over 30°). As the semi-major axes increase, the range of *i* values noticeably narrows (the inclinations approach the middle of their maximum range). Such a distribution of orbits can be explained by the orientation of the KBP spin axis at the moment of disruption.

As noted in Kazantsev (2023), the KBP spin axis at the moment of disruption must have been inclined at a small angle to the plane of the ecliptic and directed toward the Sun. It is precisely this orientation of the KBP spin axis that can explain the apparent similarity between the distributions of model orbits of the KBP fragments in the *a–e* and *a–i* coordinates and the corresponding distributions of the actual orbits of KB bodies.

According to the law of conservation of angular momentum, immediately after the breakup the directions of the spin axes of the fragments, especially the large ones, were close to the direction of the spin axis of the parent body (KBP). The KBP destruction point was also the point of intersection of the orbits of all newly formed bodies. Therefore, the spin axes of the fragments were inclined at small angles not only to the plane of the ecliptic. The main thing was that the spin axis of each body lay practically in the plane of the body's orbit. But at the moment the parent body is disrupted, a brief interaction occurs between the fragments. In this case, the spin pole of an individual body may rise above the orbital plane or fall below the plane. The spin axes of larger bodies are less displaced. The consequence of such, albeit small, displacements will be a situation where the spin axes even of the largest bodies will be inclined at small angles to the planes of their orbits, while the angular distances of the axes from the perpendiculars to the orbital planes (obliquity) may be either less or greater than 90°.

In the future, if the body does not experience any significant collisions with other bodies, ~~then~~ the orientation of its spin axis relative to the plane of its own orbit should change only due to precession under the influence of external bodies.

Such external bodies are, first and foremost, the Sun and (if any) the body's own satellites. In most cases the influence of planets and other surrounding bodies is much smaller and can be neglected.

In average, over time, Solar precession will not change the tilt of each body's spin axis relative to the plane of its orbit (hereinafter, angle $\varepsilon'$). The axis itself will trace a cone around the perpendicular to the orbital plane. Therefore, to characterize the precession of a body's spin axis, it is generally sufficient to consider the gravitational influence of its own satellites. In some cases, even for bodies with satellites, the influence of other external bodies will also play a certain role in the orientation of the axes.

## 2. Orientation of the spin axes of the dwarf planets

As noted above, the values of the angles $\varepsilon'$ of large SD bodies and bodies in the 3:2 resonance with Neptune should be relatively small. Therefore, we first consider the orientation of the spin axes of the four dwarf planets in the KB.

### 2.1 Pluto's spin axis

According to the NASA website, the equatorial coordinates of Pluto's pole are: $\alpha = 132.99°$, $\delta = -6.16°$. Then the value of $\varepsilon'$ for Pluto will be 29.5°.

Pluto has five satellites: Charon, Nix, Hydra, Kerberos, and Styx. Charon's mass is approximately 5 orders of magnitude greater than the mass of each of the other four satellites. In addition, Charon is located 2-3 times closer to Pluto than the other satellites. Therefore, only Charon can effectively influence the orientation of Pluto's axis.

Any external force $\boldsymbol{F}$ acts on a rotating body in such a way that it attempts to shift the body's principal moment of inertia vector $\boldsymbol{J}_2$ into the plane in which the vector $\boldsymbol{F}$ lies. As a rule, the vector $\boldsymbol{J}_2$ lies in or near the equatorial plane of the rotating body. Charon's orbital plane almost coincides with Pluto's equatorial plane. The angle between these planes is only 0.8°. Therefore, the spatial orientation of Pluto's spin axis should change little under Charon's influence.

It can be added that the orbital planes of Pluto’s four other satellites are also inclined to the plane of its equator by angles of less than 1°. Therefore, their (albeit slight) influence on Pluto's spin axis should qualitatively coincide with Charon's influence.

Usually, the presence of a relatively massive satellite around a planet makes it possible to practically neutralise the influence of other planets on the precession of the spin axis. Thanks to the Moon, Earth's spin axis is inclined to the perpendicular to the plane of the ecliptic at an angle of 27.5° with deviations of only 1.5°.

Charon's mass relative to Pluto's mass is significantly greater than the Moon's mass relative to Earth's mass. And the distance between Charon and Pluto is much smaller than that between the Moon and Earth. But Charon's rotation is synchronous with Pluto's rotation around its own axis. Both rotation periods are 6.387 Earth days. Therefore, the influence of the other external bodies (planets) on the precession of Pluto's spin axis can be considered as an influence on the Pluto-Charon system as a whole, i.e., as a single rigid body.

This approach had been used to estimate the precession of Pluto’s axis long before the New Horizons mission in the paper by Dobrovolskis & Harris (1983). The paper determined that the inclination of Pluto’s spin axis to the perpendicular to its orbit varies from 102° to 126° with a period of about 3 million years. That is, the inclination of Pluto’s axis to the plane of its orbit varies from 12° to 36°. Since then, this estimate has not been refined. Although four other satellites of Pluto had already been discovered, their influence on the precession of Pluto’s axis can apparently be neglected.

### 2.2 Haumea's spin axis

The dwarf planet Haumea has two satellites: Hi'iaka and Namaka. Hi'iaka is approximately 10 times more massive than Namaka. Namaka orbits Haumea approximately twice as close as Hi’iaka (Ragozzine & Brown 2009).

Unlike other dwarf planets, Haumea, in addition to its own satellites, has also a ring around it. The presence of the ring was reported in Ortiz et al. (2017) based on observations of stellar occultations on January 21, 2017, at several observatories. Analysis of the observational data made it possible to identify not only the ring itself but also its orientation in space. In Ortiz et al. (2017), the equatorial coordinates of the ring pole were determined to be $\alpha = 285.1° \pm 0.5°$, $\delta = -10.6° \pm 1.2°$. The same paper states that the coordinates of the ring pole coincide within the errors with the orbital pole of Haumea’s larger satellite, Hi’iaka's: $\alpha = 283.0° \pm 0.2°$, $\delta = -10.6° \pm 0.7°$. Thus, the authors concluded that the ring plane and Hi’iaka's orbit should lie in the plane of Haumea’s equator. The inclination of Hi’iaka's orbit to the plane of Haumea’s equator is 1°. This means that Haumea’s spin axis should, in fact, coincide with the poles of the ring and Hi’iaka’s orbit. Namaka's orbit is inclined at the angle of 12.8° to the plane of Haumea’s equator.

In Proudfoot et al. (2024), based on an analysis of Haumea's gravitational potential, the equatorial coordinates of the spin axis pole of the dwarf planet were determined: $\alpha = 282.9°$, $\delta = -9.7°$. The coordinate uncertainties were estimated to be less than 1°. At these coordinates, the angle between Haumea's spin axis and the perpendicular to its orbit is 87.1°. Therefore, Haumea's spin axis is inclined to the plane of its own orbit by only 2.9°.

The coordinates of Haumea's pole obtained in Proudfoot et al. (2024) are very close to the coordinates of the ring pole obtained in Ortiz et al. (2017). These values effectively confirm each other. That is, it can be confidently assumed that Haumea's spin axis is inclined at a small angle to the plane of its orbit.

Furthermore, in Proudfoot et al. (2024), the precession of Haumea's axis was calculated. It was found that the coordinates of Haumea's pole, in both ecliptic longitude and latitude, vary with amplitudes of less than 0.5° and a period of about 600 years. Such changes are caused by Hi’iaka's influence. Superimposed on these periodic oscillations are others, with an amplitude approximately 10 times smaller and a period 20 times shorter. These oscillations are caused by Namaka's influence.

The oscillations of Haumea's pole coordinates with a period of hundreds of years occur around a relatively constant mean value, i.e., without any systematic drift. Taking into account that solar precession does not change the inclination of Haumea’s axis to the plane of its orbit, it can be concluded that this angle must always be small.

### 2.3 Eris's spin axis

The dwarf planet Eris has one satellite, Dysnomia. The orientation of Eris's spin axis has not yet been definitively determined. However, Holler et al. (2021) found that the inclination of Dysnomia's orbital pole relative to the plane of Eris's heliocentric orbit is about 78.3° with an accuracy of 0.7°. Thus, Dysnomia's orbital pole is located approximately 12° from the plane of Eris's heliocentric orbit.

If Dysnomia's orbital plane is close to Eris's equatorial plane, then Eris's spin axis is also inclined at a small angle to the plane of its orbit. This assumption is quite logical.

Moreover, Bernstein et al. (2023) showed that Eris's axial rotation period is synchronous with Dysnomia's orbital period around Eris. Such a relationship between the periods indicates a very small inclination of Dysnomia's orbital plane relative to Eris's equatorial plane.

If Dysnomia's rotation around Eris is indubitably synchronous with Eris's axial rotation, then Dysnomia cannot stabilise the orientation of Eris's spin axis. This is precisely the situation observed with Pluto and Charon. In this case, the influence of the planets may cause additional precession of Eris's axis. However, calculations of this precession are not mandatory. For an approximate estimate, one can use the available calculations for Pluto (Dobrovolskis & Harris, 1983), for which the inclination of the spin axis to the orbital plane varies with an amplitude of 12°. For Eris, the amplitude should be smaller, since this dwarf planet orbits farther from the giant planets than Pluto does.

So, it is quite reasonable to assume that Eris's spin axis is tilted at a small angle to the plane of its orbit (about 12°). And this angle is not supposed to change much over time.

### 2.4 Makemake's spin axis

A reliable estimate of the orientation of Makemake's spin axis has not yet been obtained. Nevertheless, some approximate estimate of this orientation can be made based on the orbit of Makemake's recently discovered satellite. The existence of such a satellite was first reported by Parker et al. (2016). The conclusion was based on individual Hubble Space Telescope images from 2015. The authors suggest that the satellite had not been discovered previously due to the specific orientation of its orbit relative to Earth. The satellite's orbit may be oriented edge-on to the line of sight. With such a configuration, the satellite can only be seen at those moments when it is as far as possible from Makemake's disk, which happens quite rarely.

Ibid, Parker et al. also estimated the tilt of Makemake’s spin axis relative to the plane of the dwarf planet’s orbit. This tilt should be in the range from 12° to 44°. The conclusion was made under the assumption that the satellite's orbital plane lies in the plane of Makemake’s equator.

Later, Bamberger (2025) confirmed the presence of a satellite orbiting Makemake, based on an analysis of Hubble Space Telescope images from 2015 to 2019. According to his measurements, the satellite's orbital plane is oriented nearly edge-on to the line of sight, which is consistent with the data by Parker et al. (2016).

Thus, if the orbital plane of Makemake's satellite is inclined at a small angle to the plane of the dwarf planet's equator, then Makemake's spin axis should be inclined at a relatively small angle to the plane of its own orbit. And this orientation of the axis should change little over time.

## 3. Orientations of the spin axes of other bodies in the Kuiper belt

From the above, it can be concluded that the orientation of the spin axes of the four dwarf planets in the Kuiper Belt provides fairly strong support for the mechanism proposed in Kazantsev (2023) for the formation of the scattered disk in the KB, including dwarf planets and plutinos, as a result of the destruction of one large body. These arguments should already be sufficient to consider the proposed mechanism quite plausible. After all, it is rather difficult to explain such a specific orientation of the spin axes of the dwarf planets by other mechanisms of their formation.

However, the justification of the proposed mechanism can be continued. For this purpose, it is worth considering other bodies in the Kuiper Belt that have satellites. These should be bodies with calculated orbital elements of their satellites. To date, several dozen such bodies have been identified.

### 3.1 Approach to the selection of the SD bodies and plutinos

At the outset, it is necessary to determine which of the objects belong to the scattered disk or to the plutinos. It is precisely these group of objects that could have formed as a result of the destruction of one large body. Usually, objects on orbits with perihelion distances $q$ from 30 to 40 AU are classified as belonging to the scattered disk.

Most likely, the perihelion distances of the orbits of some scattered disk objects may slightly extend beyond these limits. However, in this paper, for the sake of sample purity, the specified range of $q$ values will be used.

In addition to the value of $q$, an important parameter for selecting scattered disk objects is the orbital inclination $i$. The reason is that objects of the cold component of the Kuiper Belt also have $q$ values from 30 to 40 AU. The cold component is thought to have an origin distinct from that of the other Kuiper Belt objects. The cold component includes objects on orbits with $a$ = 42–47 AU (Li & Chiang 2025) that have low eccentricities and inclinations.

There are different estimates of the cold component's inclination range. Petit et al. (2011) use the range $i < 5°$. Lykawka & Mukai (2007) show that the range $i < 10°$ is preferable. Meanwhile, Peixinho, Lacerda & Jewitt (2008) consider that this range should be even wider: *i* < 12°; but it is important that in that paper the distribution of KB body populations is based not only on orbital characteristics but also on the bodies' physical properties. They further show that bodies on orbits with inclinations *i* < 12° have uniformly red colours, while those on orbits with *i* > 12° are blue. This colour separation is much clearer at *i* = 12° than at *i* = 5°. Therefore, the authors conclude that bodies on orbits with *i* < 12° should be assigned to the cold component.

Given the importance of separating cold-component bodies from SD bodies, a separate study has been conducted for this paper. Using data from the Minor Planet Center (MPC) catalogue for May 2025, the distribution of KB orbits in the *a–i* coordinates was constructed. In this distribution, within the range of *a* values from 40 to 50 AU, a concentration of orbits with *a* = 42 – 47–48 AU and *i* < 5º–6º is indeed visible (Fig. 1). Purely visually, these orbits are usually attributed to the cold component.

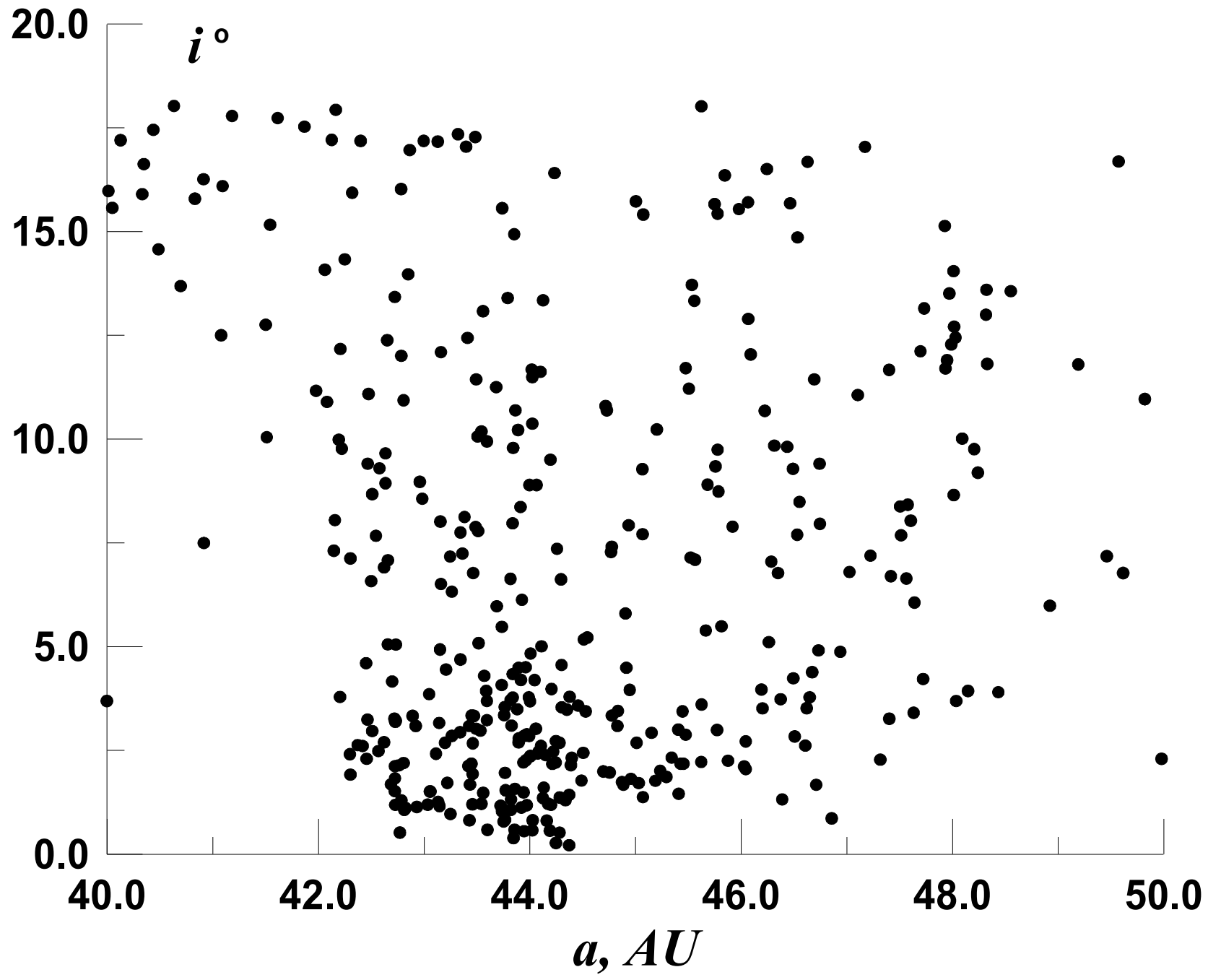

Figure 1. The *a–i* distribution of KB orbits for *a* < 50 AU

From this group, the first 60 bodies (by number) with absolute magnitudes $H < 7^m$ were selected. Numerical calculations of the evolution of their orbits were carried out for intervals of up to 8 million years. A description of the method and calculation program can be found in Kazantsev & Kazantseva (2021).

The calculations have shown that the value of *i* for the vast majority of orbits remains within 5° throughout. The inclinations of some orbits reach 7º–8°, and those of individual orbits – even up to 10°.

The range of changes in the orbital inclinations of this sample does not cover the entire possible range of changes in the *i* values of the orbits of the cold component. However, even according to the presented calculations, it can be assumed that the orbital inclinations of the cold component may noticeably exceed 5°. Some orbits may attain larger inclinations from the very beginning when the cold component population is formed. In addition, this may result from changes in orbital elements caused by external perturbations. Therefore, selecting cold-population bodies relying solely on orbital elements is not entirely correct. It is also important to consider the surface colour, which does not change with changes in the orbital elements. Taking into account the results of Peixinho et al. (2008), if the surface colour is unknown, it is better to include bodies on orbits with $i > 12^{o}$ in the SD population. Some SD body orbits may have smaller inclinations. However, in this case, the key point is that the resulting sample of SD bodies does not include bodies from the cold component.

### 3.2 Selected bodies and their characteristics

In addition to the four dwarf planets, the author found nine numbered SD and plutino bodies in published sources that have satellites with determined orbital elements. These are bodies with the following numbers in the MPC catalogue: 26308, 38628 (Huya), 55637 (Uni), 60458, 90482 (Orcus), 120347 (Salacia), 174567 (Varda) 225088, (Gonggong), 229762 (G!kun||'homdima).

In addition, another body from the Kuiper Belt was selected for further consideration, the one that belongs to neither the SD nor the plutinos. This is TNO 50000 Quaoar.

The sizes and orbital elements of these bodies are given in Table 1. The orbital elements correspond to the MPC catalogue for May 2025. The SD bodies and plutinos are listed in descending order of size.

Table 1. The sizes and orbital elements of the selected bodies

| Number (Name) | ***D***, km | ***a***, AU | ***e*** | ***i***° |
|---|---|---|---|---|
| 225088 (Gonggong) | 1230 | 66.94 | 0.502 | 30.82 |
| 90482 (Orcus) | 910 | 39.29 | 0.223 | 20.56 |
| 120347 (Salacia) | 846 | 42.18 | 0.102 | 23.93 |
| 174567 (Varda) | 720 | 45.55 | 0.144 | 21.52 |
| 55637 (Uni) | 664 | 43.00 | 0,146 | 19.40 |
| 229762 (G!kun\|\|'homdima) | 662 | 74.58 | 0.496 | 23.34 |
| 38628 (Huya) | 415 | 39.20 | 0.272 | 15.48 |
| 26308 | 270 | 47.98 | 0.371 | 13.48 |
| 60458 | 100 | 59.44 | 0.403 | 19.69 |
| 50000 (Quaoar) | 1090 | 43.16 | 0.037 | 7.99 |

Table 1 shows that 7 of the 9 bodies belong to the scattered disk. According to the semi-major axis of the orbit, 38628 (Huya) and 90482 (Orcus) should belong to the plutinos. To confirm this, numerical calculations of the evolution of these orbits were performed, accounting for perturbations from the four giant planets over intervals of up to 40000 years. The calculations showed that both orbits are indeed in the 3:2 resonance with Neptune. The semi-major axes oscillate around the mean value $a = 39.50$ AU with a period of about 10000 years.

For further analysis, it is appropriate to consider the positions of the orbits of the selected bodies against the background of the positions of the orbits of other KB bodies. The distribution is presented in Fig. 2 for the *a–e* coordinates of the orbits of KB bodies with absolute magnitudes $H < 12^{m}$ according to the MPC catalogue for May 2025.

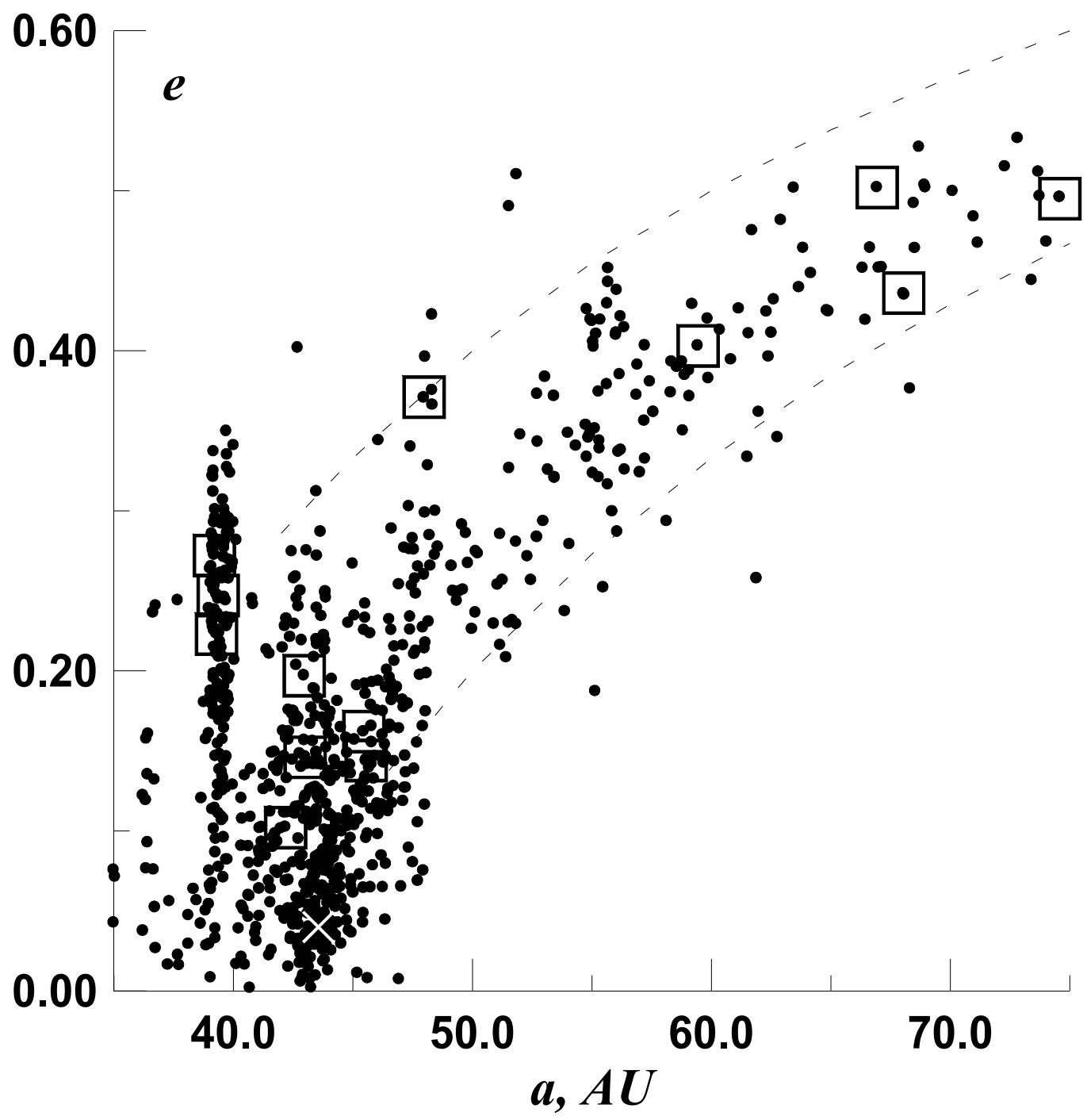

The scattered disk is bounded by the dashed lines $q$ = 30 AU and $q$ = 40 AU. The semi-major axes of the orbits of SD objects reach values of $a$ > 500 AU. Here, it suffices to show the orbits with $a$ < 75 AU. The plutinos' orbits form a vertical cluster between 39 and 40 AU.

The squares indicate the orbits of the four dwarf planets and the other selected KB objects. It is clear from the Figure that all 13 selected KB objects belong to the SD or to the plutinos.

Figure 2. The *a–e* distributions of the Kuiper belt bodies' orbits.

In Kazantsev (2023), it is shown that during the disruption of a large body, the distribution of model orbits in the *a–e* coordinates closely matches the distribution of real orbits of the scattered disk and plutinos, as shown in Fig. 2.

The clustering of orbits with $e$ < 0.1 in the interval $a$ from 40 to 50 AU represents the orbits of the cold component. The white cross marks the orbit of Quaoar.

Now it is worth considering the orientation of the spin axes of these bodies in the order given in Table 1.

**TNO 225088 Gonggong**

The size of Gonggong is about 1230 km, while its satellite, Xiangliu, is less than 100 km (Kiss et al. 2019). The authors of that paper also estimated the orientation of the satellite's orbit. The satellite's motion can be either prograde or retrograde. Kiss et al. (2019) give the coordinates of the satellite's orbital pole for both motion options. In particular, the equatorial coordinates for prograde motion are: $\alpha$ = 301,99°, $\delta$ = 6,91°, and for retrograde motion: $\alpha$ = 14,10°, $\delta$ = –29,14°. Based on these data and the known orbital elements of Gonggong, it is possible to calculate the angular distance of the satellite's orbital pole from the plane of Gonggong's orbit. For both prograde and retrograde motion, this distance is about 38° with a difference of less than 1°. The value of the angle $\varepsilon'$ should be approximately the same.

**TNO 90482 Orcus**

90482 Orcus has a satellite, Vanth. The diameters of Orcus and Vanth are 910 km and 475 km, respectively (Brown & Butler 2018).

In the paper of Grundy et al. (2019b), based on an analysis of more complete observational data for 17 KB bodies with satellites, the orientations of the satellites' orbits relative to the heliocentric orbits of the bodies themselves were updated or recalculated. In particular, for Orcus, the angle between its orbit and Vanth's is 106.7°. That is, the pole of the satellite's orbit is 16.7° from the plane of Orcus' orbit. That paper states that the uncertainty in the measured angle between the orbits is less than 0.5°. The inclination of Orcus's spin axis to the plane of its orbit should also be about 17°.

**TNO 120437 Salacia**

Salacia has a satellite, Actaea. Grundy et al. (2019b) calculated the mean sizes of these bodies. They are 846 km and 284 km, respectively. That paper also calculated the angle between the orbital planes of Actaea and Salacia. The calculated value is 41.4°. In this case, the angular distance of the pole of Actaea's orbit from the orbital plane of Salacia will be 48.6°.

Later, Collyer et al. (2025) analysed the Salacia–Actaea system using new observational data. They concluded that the two bodies are fully synchronous in their rotation. This means that Actaea's orbital plane must lie very close to Salacia's equatorial plane. In addition, that paper provides the equatorial coordinates of the pole of Actaea's orbit. Within the uncertainty (< 1.5°), these coordinates are consistent with the data in Grundy et al (2019b). Thus, it can be assumed that the value of the angle ε′ for Salacia should be about 49°.

**TNO 174567 Varda**

Varda has a satellite, Ilmarë. Their sizes differ somewhat across sources, though not significantly. In particular, Grundy et al. (2015) estimate the sizes of Varda and Ilmarë at approximately 700–740 km and 280–380 km, respectively. The range of sizes is due to the possible range of the bodies' albedos.

Additionally, Grundy et al. (2015) calculated the coordinates of Ilmarë's orbital pole. However, due to uncertainty in the direction of the satellite's motion, the authors present two variants of the pole coordinates: both for prograde and retrograde motion. The equatorial coordinates for one direction: $\alpha = 273°$, $\delta = -11°$; for the other one: $\alpha = 229.5°$, $\delta = 4.9°$.

Based on these data and Varda's orbital elements, the angular distance of Ilmarë's orbital pole from Varda's orbital plane can be calculated. For the first case, this distance is 9°, and for the second one, 7.4°. That is, in both cases, these values are quite close. Therefore, an average value of 8° can be taken (with an accuracy of about 1°). It can be assumed that for Varda the angle ε′ is also about 8°.

**TNO 55637 Uni**

All the necessary data for this binary system are given in Brown (2013). The sizes of Uni and its satellite are estimated at approximately 664 km and 190 km, respectively. The angle between the orbital planes of Uni and its satellite is estimated at 65°. The errors in the angular orbital elements of the satellite do not exceed 1°. In this case, for Uni the angle ε′ should be about 25°.

**TNO 229762 G!kunǁ'homdim**

All the necessary data for this binary system can be found in Grundy et al. (2019a). The sizes of the body and the satellite are circa 632 km and 142 km, respectively. The angle between the satellite's orbit and the body's heliocentric orbit is estimated at 32.3° with an uncertainty of 0.6°. Then, for this TNO the angle ε′ is approximately 58°.

**TNO 38628 Huya**

Huya has a mean diameter of about 415 km, and its satellite has a diameter of 165–243 km (Rommel et al., 2025). That paper obtained the equatorial coordinates of the satellite's orbital pole: α = 20.8° ± 2°, δ = 34.9 ° ± 2°. In this case, the satellite's orbital pole is located at a distance of 34.2° from the plane of Huya's heliocentric orbit. The angle ε′ for Huya should be approximately the same.

**TNO 26308**

Data on the sizes of the body and its satellite can be found at https://www.johnstonsarchive.net/astro/tnoslist.html. Their sizes are given as 270 km and 81 km, respectively. In Grundy et al. (2019b), the angle between the satellite's orbit and the body's heliocentric orbit was given as 75.5°. Therefore, the angle ε′ for this TNO should be approximately 15°.

**TNO 60458**

The sizes of this TNO and its satellite can also be found on the website: https://www.johnstonsarchive.net/astro/tnoslist.html. Their sizes are listed as 100 km and 77 km, respectively. Also, in Grundy et al. (2019b), the angle between the satellite's orbit and the body's heliocentric orbit was calculated to be 55.5°, with an uncertainty of about 2°. Then the angle ε′ for the body should be about 35°.

**TNO 50000 Quaoar**

Quaoar belongs to the cold component. And this is not only because the inclination of its orbit lies within the range of orbital inclinations of some bodies of the cold component. The main thing is the colour of Quaoar's surface.

Brown (2011) reported that Quaoar appears red, which may be due to the presence of methane on its surface. A similar conclusion was reached by Barucci et al. (2015). Therefore, based on both inclination and colour, Quaoar can confidently be classified as a body of the cold component in the KB. Nevertheless, it is worth assessing the orientation of Quaoar's spin axis.

Its size is about 1090 (Kiss et al. 2024). Quaoar has a satellite, Weywot, about 200 km in size. In addition, two ring systems were discovered around Quaoar. Proudfoot et al. (2025) determined the equatorial coordinates of the pole of Quaoar's rings: α = 259.5°, δ = +55.0°. The uncertainties of the angles are less than 0.5°.

Using these values, the angular distance of the ring pole from the plane of Quaoar's heliocentric orbit can be calculated. It is about 70.5°. That is, the value of the angle ε′ for Quaoar should be approximately the same.

## 4. Summary data on the orientation of the spin axes of the KB bodies

Now, let us consider all the data obtained above on the orientation of the spin axes of the 13 bodies in the Solar System (Table 2). The table lists the names of the bodies, their sizes, the angles ε′ for the bodies, and their accuracy. For Makemake, the mean value of the specified range of possible values of ε′ (12° – 44°) is given.

Table 2. The sizes and the ε′ values for the KB bodies

| Name (Number) | ***D***, km | ε′° | Δε′° |
|---|---|---|---|
| Pluto | 2377 | 29.5 | 0.5 |
| Haumea | 1560 | 2.9 | 1.0 |
| Eris | 2326 | 12 | 3.0 |
| Makemake | 1430 | 28 | 18.0 |
| Gonggong | 1230 | 38 | 3.0 |
| Orcus | 910 | 17 | 2.5 |
| Salacia | 846 | 49 | 3.5 |
| Varda | 720 | 8 | 3.0 |
| Uni | 664 | 25 | 3.0 |
| G!kun\|\|'homdima | 662 | 58 | 3.0 |
| Huya | 415 | 34 | 4.0 |
| 26308 | 270 | 15 | 5.0 |
| 60458 | 100 | 35 | 4.0 |
| Quaoar | 1090 | 70.5 | 2.5 |

An important indicator should be the accuracy of the angles ε′. For Pluto and Haumea, the pole coordinates, and hence the values of the angles ε′, are determined directly. In Table 2, their uncertainties are given as 0.5° and 1°, respectively. Although for Pluto, the uncertainty is probably even smaller.

For the other bodies, the angles between the orbital planes of the body and its satellite are defined in the aforementioned publications. The values of the angles ε′ for these bodies were obtained under the assumption that the satellite's orbital plane lies close to the equatorial plane of the corresponding body.

Such an approach introduces an additional error into the value of ε′. What could the magnitude of this error be? In determining the angles ε′, data on the orbits of satellites orbiting bodies whose sizes are on the order of tens of per cent of the bodies' sizes were used. Only in Gonggong does the satellite's size amount to 8% of the body's size. And the orbital planes of such satellites should lie close to the equatorial planes of the central bodies. As is the case for the planets in the outer part of the Solar System, as well as for Pluto and Haumea. In addition, the synchronous rotation of Eris and Salacia with their satellites indicates that the orbital planes of these satellites are close to the bodies' equatorial planes.

The inclinations of the orbital planes of the major satellites of Pluto and Haumea to the equatorial planes of the bodies do not exceed 1°. In view of this, it was assumed that for other bodies, the angles between the orbital planes of the satellites and the equators of the bodies do not exceed 2°. Therefore, the additional uncertainties in the values of the angles ε′ are also 2°. The uncertainties in the values ε′ in Table 2 are the uncertainties in determining the angles between the orbital planes of the bodies and their satellites, increased by 2°.

A separate case is body 26308. In the paper by Grundy et al. (2019b), the angle between the orbital planes of the body and its satellite is given, but without an uncertainty estimate. Among the other bodies in Table 2, the maximum value of such an uncertainty is 2°. It was assumed that for body 26308, the uncertainty is 3°.

For clarity, the dependence of the angle ε′ on the body size was plotted using the data from Table 2 (Fig. 3).

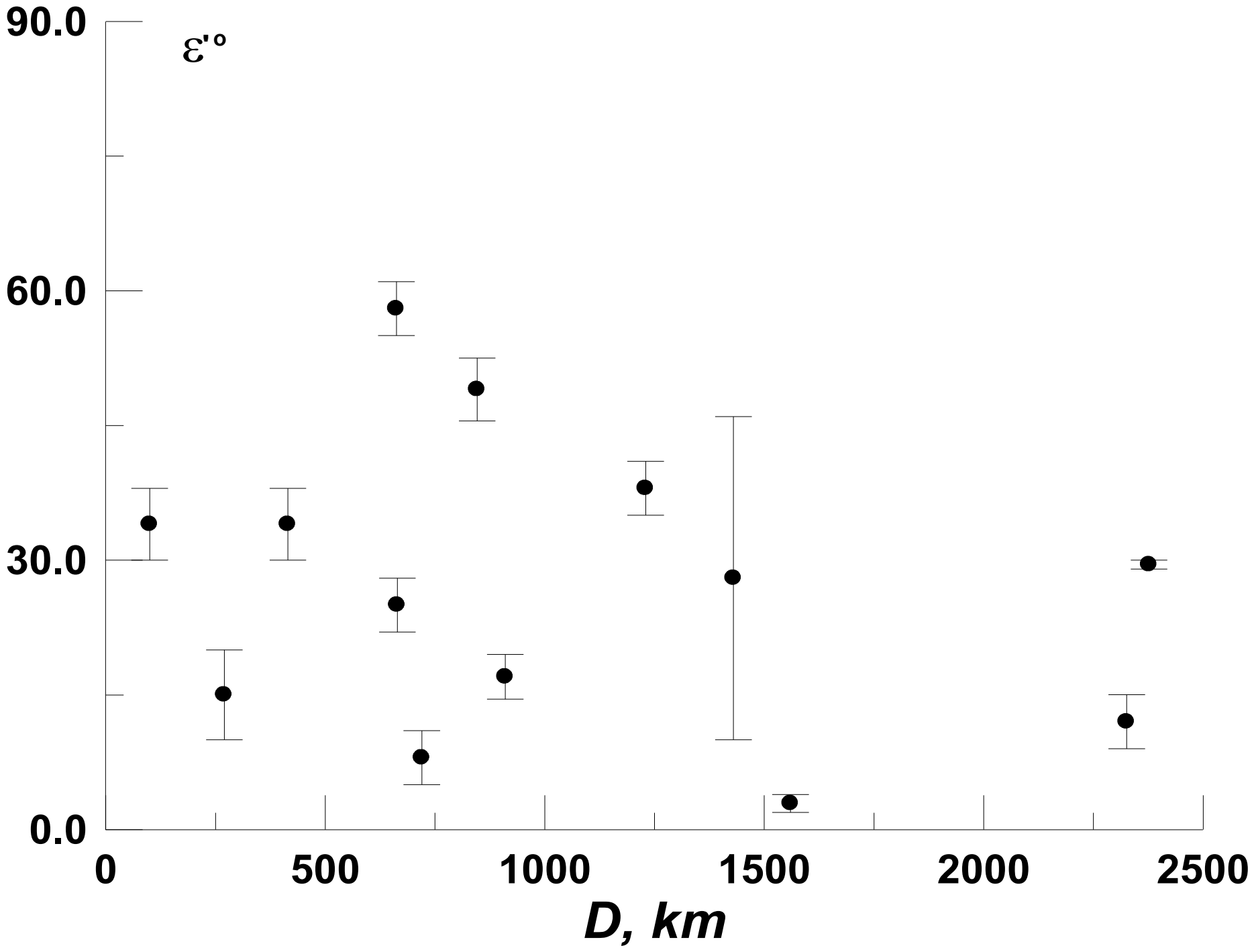

Figure 3. The ε′(*D*) dependence for the KB bodies.

According to the previously proposed mechanism for the formation of scattered disk bodies, the spin axes of all large bodies should be inclined at relatively small angles to the planes of their orbits. It is difficult to say accurately what these angles should be. Moreover, some angles may have changed as result of precession after the parent body is destroyed.

At the moment of KBP disruption, the spin axes of the largest fragments should have been shifted by minimal angles relative to the spin axis of the parent body. Dwarf planets can be included among such bodies.

The effect of precession should be greatest for the orbits of plutinos that are located closest to the giant planets. As a result of the precession, Pluto's spin axis changes within the range of 12° – 36°. The effect of the precession on other dwarf planets should be smaller. Therefore, the value of the angle ε′ for dwarf planets should be greatest for Pluto. And that is indeed the case. Although, given the margin of error, the value of Makemake's angle ε′ may be even larger. But, firstly, there is not yet enough data for this dwarf planet to obtain a more precise value ε′. Secondly, even the maximum possible value of ε′ (46°) would not contradict the proposed mechanism for the origin of SD bodies. After all, Makemake is the smallest among the dwarf planets.

The spin axes of smaller bodies may be more displaced relative to the spin axis of the parent body. First, at the moment the parent body breaks apart, during the brief interaction between the fragments, the spin axes of the smaller bodies may shift more. Second, after the fragments have separated, meteorites can sometimes fall onto the surfaces of the newly formed bodies. This can also alter the orientation of the bodies' spin axes. Such a process should be more noticeable for smaller bodies.

Therefore, it is quite logical to assume that as the sizes of bodies decrease, the range of ε′ values should increase. At the same time, it is not necessarily the case that smaller bodies should have larger ε′ values. Some relatively small bodies may disperse without significantly shifting their spin axes. The main point is that for smaller bodies, the range of ε′ values should be wider. The $\varepsilon'(D)$ dependence confirms this assumption in Fig. 3.

## 5. Estimates of the distributions of angles ε′ using a statistical test

In addition to the above-mentioned arguments in favour of a relatively small range of values of the angle ε′ of large SD bodies, a formal statistical test can be applied to assess the significance of such specific distributions of angles ε′.

In this case, it is convenient to use the well-known Kolmogorov test, a detailed description of which can be found elsewhere, in particular, in Müller, Neumann & Storm (1973), Chapter 19.

The general population should be considered as a set of values of sin(ε′). Under isotropic distribution of the angles ε′, the values of sin(ε′) are uniformly distributed on [0, 1]. This population will have a continuous distribution function $F_0(\sin(\varepsilon')) = \sin(\varepsilon')$.

The corresponding empirical function $F_n(\sin(\varepsilon'))$, constructed from a sample of $n$ values of the $\sin(\varepsilon'(i))$ ($i = 1 \div n$), will be stepwise. On the interval $\varepsilon' \leq \varepsilon'(1)$, $F_n(\sin(\varepsilon'))$ will equal 0. On each subsequent interval $\varepsilon'(i) < \varepsilon' \leq \varepsilon'(i+1)$ (for $i < n$), $F_n(\sin(\varepsilon'))$ will equal $i/n$. For $\varepsilon' > \varepsilon'(n)$, $F_n(\sin(\varepsilon'))$ will equal 1.0.

The conformity of the empirical function $F_n(\sin(\varepsilon'))$ to the isotropic distribution of ε′ values (the null hypothesis) is determined by the value of the statistic $D_n$. As the ε′ values are small, the one-sided test one can use. At that the statistic $D^+_n$ is the maximum value of the difference between $F_n(\sin(\varepsilon'))$ and $F_0(\sin(\varepsilon'))$ at equal values of ε′.

For a sample size $n$ and significance level α, there is a corresponding critical value $K_{n;\alpha}$. If $D^+_n > K_{n;\alpha}$, then the null hypothesis is rejected at the significance level α. It is considered that a significance level of 0.05 is sufficient to reject the null hypothesis.

The value of the $D^+_n$ statistic can be derived quite easily from the data in Table 2. The necessary critical values $K_{n;\alpha}$ can be found in Table 19A of Müller et al. (1973).

First, an estimate can be made for the four dwarf planets. The maximum ε′ value of these bodies is in Pluto (29.5°). Therefore, already at ε′ > 29.5° the value of the corresponding empirical function $F_4(\sin(\varepsilon'))$ will be equal to 1. The value of the function $F_0(\sin(\varepsilon'))$ at ε′ = 29.5° is 0.492. Thus, the value of the statistic $D^+_4$ will be 1 – 0.492 = 0.508.

The critical value $K_{n;\alpha}$ for $n$ = 4 and significance level α = 0.05 is 0.624 i.e. greater than $D^+_4$.

Therefore, the assumption of isotropic distribution of the angles ε′ for dwarf planets cannot be rejected at the 0.05 significance level, i.e. with sufficient justification.

Of course, a sample of 4 bodies is too small. Therefore, it is worth assessing the significance for a larger sample of bodies from Table 2, for example, those with $D$ > 700 km. Now the value of $n$ = 8. The maximum value of the difference between $F_n$ and $F_0$ in this case is at the ε′ value of Gonggong ($i$ = 7, ε′ = 38°).

Then the $D^+_8$ = 7/8 – sin(38°) = 0.259. The critical value $K_{n;\alpha}$ for $n$ = 8 and the significance level of α = 0.05 is 0.454. That is, the assumption of isotropic distribution of the angles ε′ for bodies with sizes $D$ > 700 km cannot be rejected as well.

A similar assessment for all 13 bodies from Table 2 can be omitted. Among them are bodies with significantly smaller sizes (up to 100 km in diameter). It is clear that, among SD bodies and plutinos, the number of bodies of this size is much greater. But to date, the orientation of the spin axes has been estimated for a relatively small fraction of these bodies. Therefore, the statistical distribution of the angles ε′ of these 13 bodies will not correspond to the distribution for all existing bodies with sizes $D$ > 100 km. Moreover, as the sizes decrease, the range of ε′ values should increase.

## 6. Discussion

The application of the statistical test did not refute the assumption of an isotropic distribution of the angles ε′ of SD bodies and plutinos. However, it does not reject the proposed mechanism for the origin of these bodies. First, immediately after the breakup of the KBP, the range of ε′ values for the newly formed bodies could have been narrower than it is today.

Secondly, one should take into account the arguments presented above in favour of the proposed mechanism.

After all, the range of the angles ε′ for the SD bodies and plutinos is relatively small. Particularly important are the relatively small values of ε′ for all four dwarf planets. A large value of this angle for at least one dwarf planet would, in fact, contradict the reality of the proposed mechanism. The small ε′ values obtained are the main difference between this mechanism and the other existing mechanisms for the formation of both dwarf planets and the scattered disk as a whole.

A significant fact supporting the mechanism discussed here is the relatively small values of ε′ for bodies in the 3:2 resonance with Neptune: Pluto, Orcus, and Salacia. According to this mechanism, such bodies formed simultaneously with the other bodies of the scattered disk.

Moreover, it is very important that the colours of the SD and plutino bodies are similar, whereas the colours of the cold-component bodies differ noticeably (Sheppard, 2010).

Besides, in Kazantsev (2023), it is shown that during the disruption of a large body, the distributions of model orbits in the *a–e* and *a–i* coordinates closely match the distributions of real orbits of the scattered disk and plutinos.

Let us consider once again the large value of ε′ for Quaoar, which belongs to the cold component. And the point is not only that this example does not contradict the mechanism of the origin of KB bodies considered herein. As shown by Peixinho et al. (2008), SD and cold-component bodies differ in both orbital inclinations and surface colours. It can now be assumed that the values of the angle ε′ can also distinguish such bodies. And such a difference may also indicate to the mechanism of origin for each group.

The large value of the angle ε′ is also consistent with the possibility that bodies of the cold component were formed as a result of the disruption of a single large body. The orbits of this population have relatively small inclinations. This would be expected if the spin axis of their parent body was highly inclined to the plane of the ecliptic. In this case, the spin axes of the large fragments of this body will also have a similar orientation, as Quaoar does. However, this is only an assumption, the verification of which requires a separate study. Nevertheless, the significant difference between the values of ε′ for the large body of the cold component and the corresponding values for the large bodies of the scattered disk is an interesting fact.

In Kazantsev (2023), only Pluto and Haumea were considered when analysing the orientation of the spin axes of KB bodies. In this paper, later publications were used to substantiate the proposed mechanism for the origin of such bodies. This made it possible to include not only all four dwarf planets but also several other KB bodies in this argument. In addition, the effect of precession on the changes in the orientation of the spin axes was taken into account, a point not addressed in Kazantsev (2023). Overall, the new arguments presented here, together with the previous ones (Kazantsev, 2023), point to a possibility that the scattered disk and plutino objects formed as a result of the destruction of a single large body. In any case, the proposed mechanism for the origin of the scattered disk and plutino objects deserves further investigation.

## 7. Conclusions

The orientation of the spin axes of large scattered disk objects in the Kuiper Belt, including four dwarf planets, indicates that the previously proposed mechanism of their formation as a result of the disruption of a single large body is possible.

An important argument in support of such a mechanism is the orientation of the spin axes of bodies in the 3:2 resonance with Neptune. Indeed, according to the proposed mechanism, such bodies formed simultaneously with bodies of the scattered disk.

The orientation of the spin axes of the large bodies in the cold component may be an additional argument in favour of their origin, distinct from that of scattered disk bodies. In addition, such an orientation may indicate the mechanism by which these bodies originated.

## DATA AVAILABILITY

The data underlying this article are available in the Minor Planet Center catalogue and can be accessed with https://minorplanetcenter.net//

## Acknowledgements

The author is sincerely grateful to the Service for Information Monitoring and Support of Scientific and Publishing Activities of the M. Maksymovych Scientific Library of the Taras Shevchenko National University of Kyiv for providing the texts of papers that were not in the public domain.